\documentclass[twocolumn,showpacs,amsmath,amssymb,prl]{revtex4}

\usepackage{graphicx}
\usepackage{psfrag}
\usepackage{dcolumn}
\usepackage{bm}

\begin{document}


\title{The Statistics of Crumpled Paper}

\author{Eric Sultan}
\email{eric.sultan@lps.ens.fr}
 \author{Arezki Boudaoud}
\affiliation{%
Laboratoire de Physique Statistique, UMR 8550 du CNRS/Paris VI/Paris VII \\
\'Ecole normale sup\'erieure, 24 rue Lhomond, F-75231 Paris Cedex 05, France}%

\date{\today}

\begin{abstract}
A statistical study of crumpled paper is allowed by a minimal 1D model: a self-avoiding line bent at sharp angles -- in which resides the elastic energy --  put in a confining potential. Many independent equilibrium configurations are generated numerically and their properties are investigated. At small confinement, the distribution of segment lengths is log-normal in agreement with previous predictions and experiments. At high confinement, the system approaches a jammed state with a critical behavior, whereas the length distribution follows a Gamma law which parameter is predicted as a function of the number of layers in the system.

\end{abstract}

               
\maketitle

When a sheet of paper is crushed, the formation of a network of ridges is observed. The process is irreversible as energy accumulates in small regions leading to localised plastic flow. Many fundamental questions arise. What is the resistance of the crumpled ball to mechanical forces? What are the mechanisms of the cascade of energy to small scales? What is the probability distribution of lengths and energies? The two latter are similar to the questions central to hydrodynamic turbulence~\cite{frisch}. Early experimental studies~\cite{gomes87a,gomes87b,gomes90} of crumpled paper balls focused on fractal properties such as the scaling of the radius with the size of the flat sheet. The same property was used to characterise the phases of microscopic membranes~\cite{crumpling} -- such as red blood cells or graphitic oxide -- for which thermal fluctuations are relevant; these microscopic membranes raise a number of numerical and theoretical difficulties, particularly when self-avoidance is implemented~\cite{crumpling2}. Crumpled paper was also viewed as a self-affine surface which roughness was predicted and measured~\cite{tzschichholz,ploura}.

From elasticity theory~\cite{landau}, we know that thin elastic plates have two modes of deformation: bending -- which involves curving the plate -- and stretching -- which changes the distances on the plate. Bending is much less expensive energetically than stretching so that pure bending deformations should be always preferred. However, this is not possible for a number of boundary conditions as shown in~\cite{benamar}; this leads to the formation of a near-singular network of lines (stretching ridges) and points (developable cones) where the expensive stretching energy is localised. Even if  isolated ridges~\cite{witten93,lobkovsky95,lobkovsky} or d-cones~\cite{benamar,cerda99,mora,liang} are rather well-characterised, the understanding of  a full network is far from being complete. Most experimental and theoretical studies tackled situations with a small number of singularities~\cite{chaieb97,lobko2,pauchard98,cerda99,arezki_pedro} or a highly symmetric network~\cite{pogorelov,pomeau97}.

Our aim is to investigate numerically and theoretically the statistics of crumpled paper. The strength of crumpled sheets was measured and found to involve logarithmic relaxation and a critical behavior close to a compact configuration~\cite{matan}.  
Experiments also showed a broad distribution for the length of ridges, either indirectly -- through noise emission~\cite{noise,houle} -- or directly -- through geometrical measurements~\cite{blair}. This could be explained~\cite{witten} in terms of the hierarchical random breaking of ridges into smaller ones. However, full numerical approaches~\cite{lobkovsky95,patricio97,kramer97,astrom} are impeded by the complexity of the problem and did not allow statistics on crumpled paper. Here we introduce a minimal 1D model. We generate numerically a large number of equilibrium configurations and study the resulting mechanical and geometrical properties. In particular, we show a transition of the length distribution from log-normal to Gamma and we predict the parameter of the Gamma distribution using arguments analogous to those used for mixing~\cite{mixing} and spray formation~\cite{spray}.

\begin{figure}
    \centering
    \includegraphics[width=.65\columnwidth]{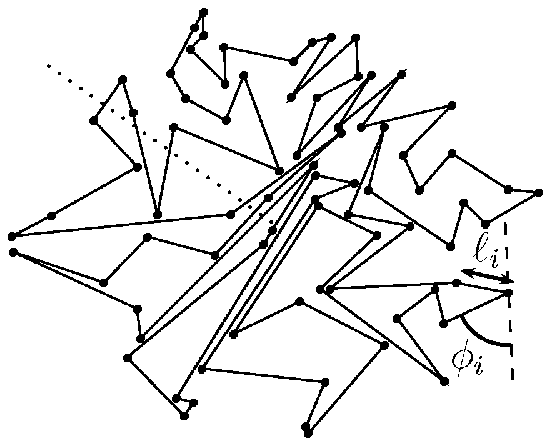}
    \includegraphics[width=.6\columnwidth]{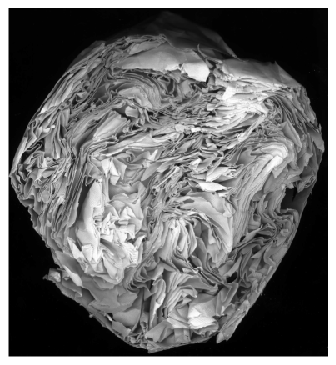}
    \caption{\textbf{a} A compact equilibrium state of the 1D model. A self-avoiding line bent at sharp angles $\phi_i$ with $N$ segments of free lengths $\ell_i$ is put in a confining quadratic potential of strength $\lambda$. Here the number of vertices is $N=80$, the thickness $h=10^{-3}$ (in units of total line length) and the confinement $\lambda=10^6$. \textbf{b}  Picture of a cut through a crumpled ball of paper, courtesy E. Couturier.}
    \label{fig1}
\end{figure}

A 2D crumpled sheet can be considered as a self-avoiding polyhedron which energy is concentrated at the edges (stretching ridges~\cite{lobkovsky}) and vertices (d-cones~\cite{benamar}). The 1D equivalent that we introduce is a self-avoiding closed line bent at sharp angles and which energy is concentrated at the vertices (Fig.~1a). The important feature here is that the position of the vertices along the line is free. This model can be viewed either as a representation of a cut through a crumpled ball of paper (Fig.~1b) or as accounting for a crumpled sheet such that all ridges are parallel and have the same length $L$. The state of the system is defined by the coordinates $(x'_i,y'_i)$ of the $N$ vertices. The total length of the line is conserved and set to 1 by defining its energy as a function of the rescaled coordinates $(x_i,y_i)=(x'_i,y'_i)/l$, $l=\sum_{i} [(x'_{i+1}-x'_i)^2+(y'_{i+1}-y'_i)^2]^{1/2}$. In order to keep the features of 2D sheets, the energy of each vertex is taken as the energy of a ridge (an edge) bent with the same angle $\phi_i$ -- defined so that $\phi_i=0$ when the sheet is flat -- and of length $L$ (the value of $L$ is unimportant as it will be scaled out). The line is put in a confining quadratic potential of strength $\lambda$. The self-avoidance is implemented through a hard-core interaction $\mathcal{E}_{\mathrm{hc}}$ with infinite cost (a large number in the simulation) between all pairs of  segments, taking into account their thickness $h$. The total energy reads
\begin{equation}
\mathcal{E}=\Sigma_i\tan^{7/3}\phi_i + \lambda \int \mathbf{r}^2(s) \mathrm{d}s+\mathcal{E}_{\mathrm{hc}}.
\end{equation}
$\mathbf{r}(s)$ is the (rescaled) position vector as a function of the curvilinear coordinate along the line. The unit of energy is the bending modulus of the sheet multiplied by the factor $(L/h)^{1/3}$ that comes from the ridge energy~\cite{lobkovsky}; the unit of length is the total length of the line. The number of vertices $N$ and the thickness $h$ -- which appears only in the hard-core interaction -- are left as parameters whereas the confinement $\lambda$ is the control parameter. This system has a large number of energy minima -- which is favored by  self-avoidance -- and our aim is to investigate the statistical properties of the corresponding equilibrium configurations. Note that we do not account for the plastic flow which reduces the stiffness of the ball and introduces a history dependance in the equilibrium states.

In the numerical minimisation of the energy, Powell's algorithm~\cite{nr} appeared as the most convenient to cope with the discontinuities of the hard-core interaction. This algorithm is built upon a set of $M$ directions ($M$ being the number of degrees of freedom) which are used to determine the directions for the successive 1D minimisations. In order to obtain independent samples of the metastable states of the system, the set of directions is chosen at random. A better convergence was obtained by running Powell's algorithm 4 times and initialising the set of directions before each run.
Care was taken in optimising the evaluation of the hard-core interaction which is expensive (typically 90\% of CPU time) because it involves all pairs of segments.
The results reported here correspond to  4 values of $N$ $\times$ 6 values of the thickness $h$ (in the range $10^{-5}$ to $10^{-2}$) $\times$ 150 samples $\times$ 40 values of the confinement $\sim 10^5$ minimisations with 100 to 200 degrees of freedom. 

We first measured the gyration radius $R_\mathrm{g}=\langle\mathbf{r}^2\rangle^{1/2}$ averaged over all samples as a function of the confinement $\lambda$ (e.g. Fig.~2 for $N=80$). For a given sample, the gyration radius depends on the choice of the sequence of applied confinements $\lambda=\lambda_1,\ \lambda_2\dots$ as in the experiments of~\cite{matan}. This can be ascribed to the large number of metastable states so that the equilibrium configuration is selected by the history of the system. The process of averaging over samples almost suppresses this dependence. At low lambda, the equilibrium state is a regular closed polyhedron so $2\pi R_\mathrm{g}\simeq 1$. When $\lambda$ reaches a threshold $\lambda_c \sim 1$, the polyhedron buckles and starts to fold. When $\lambda\sim100$, many contact points have formed. Eventually $R_\mathrm{g}$ decreases slowly with $\lambda$ to a minimal value $R_\mathrm{g}^\mathrm{j}$ value such that the system is jammed; we roughly find 
\begin{equation}
R_\mathrm{g}-R_\mathrm{g}^\mathrm{j} \sim \lambda^{-\alpha},
\end{equation}
with $\alpha=0.3 \pm 0.1$.
In this large $\lambda$ regime, we also found the total energy $\mathcal{E}$ and the energy of the ridges $\mathcal{E}_\mathrm{r}$ to scale as
\begin{equation}
\mathcal{E}\sim \lambda^\beta,\quad \mathcal{E}_\mathrm{r}\sim \lambda^{\beta_\mathrm{r}},
\end{equation}
with $\beta=1\pm 0.05$ and $\beta_\mathrm{r}=0.45\pm 0.05$. The exponents can be understood as follows.
The system reaches a jammed state of radius $R_\mathrm{g}^\mathrm{j}$, so that the main contribution to its energy is the confining potential $\sim\lambda {R_\mathrm{g}^\mathrm{j}}^2$ and $\beta=1$. The main contribution of ridges comes from U-turns, i.e. from the few ridges with angles $\phi_i=\pi/2-\epsilon$, $\epsilon \ll 1$ being a small angle. The radius of gyration approaches its limiting value as $R_\mathrm{g}-R_\mathrm{g}^\mathrm{j} \sim   \epsilon^2$. Balancing the energy of ridges $\epsilon^{-7/3}$ with the correction to the confining potential $\sim\lambda  \epsilon^2$ yields
$\alpha=6/13\simeq 0.46$ and $\beta_\mathrm{r}=7/13\simeq 0.54$. These exponents are slightly larger than in the numerics. The value of $\alpha$ differs from the experimental~\cite{matan} and numerical~\cite{astrom} value $\alpha=0.54$ for 2D sheets probably because of the different dimensionality.

\begin{figure}
    \centering
    \includegraphics[width=.8\columnwidth]{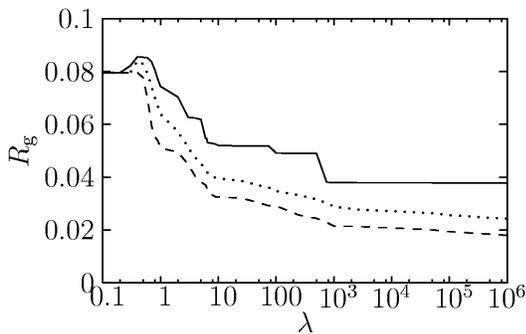}
    \caption{Gyration radius of the line $R_\mathrm{g}$ as a function of the strength of the confining potential $\lambda$. Continuous line: average over 60 samples; dotted (resp. dashed) line: maximum (resp. minimum) value of $R_\mathrm{g}$ among all samples. $N=80$ and $h=10^{-3}$. The differences between maximum and minimum are relatively large; nevertheless, the standard deviation is only of the order of $5\,10^{-4}$.} 
    \label{fig2}
\end{figure}

We examined the correlations of the normals along the line and we found them to be small except at high confinement or for zero thickness as some pairs of segments in contact are exactly aligned. The probabilty distribution of angles is symmetric with respect to zero and almost flat -- of course this probability vanishes for the maximal angles $\phi=\pm\pi$.  In contrast, the distribution of distances between two consecutive vertices appeared to be broad. We first checked that a threshold on length (up to $10\ h$) or angle (up to $10^\circ$) --  i.e. suppressing from the statistics small segments or vertices with a small angle -- does not affect the results.

At small confinement ($1<\lambda\lesssim10^4$ or $1 > 2\pi R_\mathrm{g} \gtrsim 0.2$), we observe sequential buckling events such that a part of the line folds inward reducing the radius of the ball. As predicted in~\cite{witten} and found experimentally in~\cite{blair}, the hierarchical splitting of a line leads to a log-normal distribution (see Fig.~3a) with probability density
\begin{equation}
P_\mathrm{LN}(x=\ell/\langle\ell\rangle)=\frac{1}{\sqrt{\pi} \sigma x} \exp\left(-\frac{\ln^2 x }{\sigma^2}+\frac{1}{4}\right).
\label{pln}
\end{equation}
This distribution is in agreement with our data (except for sizes comparable to the thickness $h$), with a width $\sigma$ in the range 1.0--1.8, comparable to the experimental values in the range 1.2--1.4~\cite{blair}. 
Note that these experiments were performed at low confinement -- $2\pi  R_\mathrm{g}  \gtrsim 0.6$ in units of sheet length.

\begin{figure}
    \centering
    \includegraphics[width=.85\columnwidth]{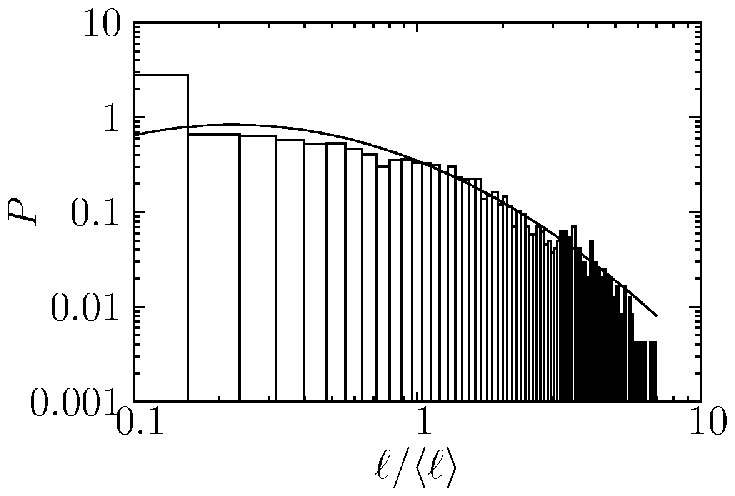}
    \includegraphics[width=.8\columnwidth]{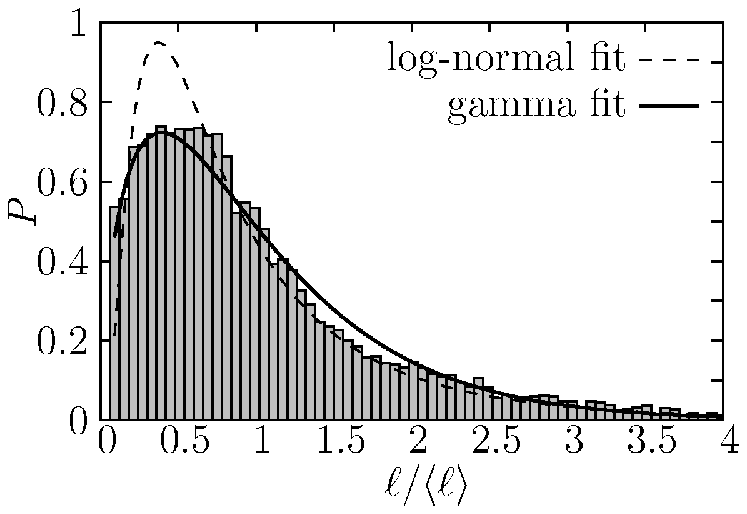}
    \caption{The distribution of segments lengths $P(\ell/\langle\ell\rangle)$ normalised by the mean length ($N=60$ and $h=10^{-3}$). \textbf{a} Low confinement $\lambda=900$. The line is a best fit to a log-normal distribution $P_\mathrm{LN}$ (Eq.~\ref{pln}) with width $\sigma=1.59\pm0.03$. \textbf{b} High confinement $\lambda=10^{6}$. The two lines ares a best fits to a log-normal distribution $P_\mathrm{LN}$ (Eq.~\ref{pln}) and to a Gamma distribution $P_\Gamma$ (Eq.~\ref{pg}) with parameter $\alpha=1.64\pm0.03$.}
    \label{fig3}
\end{figure}

In contrast, at high confinement ($\lambda\gtrsim10^4$ or $2\pi R_\mathrm{g}\lesssim 0.2$), we found that a Gamma distribution with density
\begin{equation}
P_\Gamma(x=\ell/\langle\ell\rangle)=\frac{\left(\alpha x\right)^\alpha}{\Gamma(\alpha)x}\exp(-\alpha x)
\label{pg}
\end{equation} 
provides  a better fit to the data than a log-normal distribution which fails for the most likely sizes (Fig.~3b). This can be accounted for as follows. The buckling process is no longer hierarchical which at first sight should lead to an exponential distribution of lengths $P_\mathrm{E}(x=\ell/\langle\ell\rangle)=\exp(-x)$ as the result of the random splitting of a line~\cite{feller}. However, at large confinement, many self-contact points exist and segments in contact tend to correlate. Let us define the number of layers $N_\mathrm{L}$  as the mean number of intersections of the system with a straight half-line starting from the center of the potential (see Fig.~1), averaged over all directions and all samples. $N_\mathrm{L}$ varies from 1 at low confinement to about 5 at the largest confinements (except in the case $h=0$: $N_\mathrm{L}$ can be as large as 7.5). Note that the number of layers is always smaller than the simple estimate $1/(2\pi R_\mathrm{g})$ (in the range 1--8) which reflects the "roughness" of the ball. The system can be viewed as $N_\mathrm{L}$ interacting layers. If they were independent, the distribution of lengths would be exponential $P_\mathrm{E}$. Here the lengths of segments in contact tend to average. So the the length should be the mean of $N_\mathrm{L}$ exponential random variables, i.e. a random variable with a Gamma distribution $P_\Gamma$ of parameter $\alpha=N_\mathrm{L}$. We measured the parameter from the fit to the numerical distributions and we found that, as soon as $N_\mathrm{L}\gtrsim 2.5$,
\begin{equation}
\alpha=a \ N_\mathrm{L}-b
\end{equation}
with $a=0.95 \pm 0.1$ and $b=2.05\pm 0.2$ (Fig.~4). This in agreement with our argument except that the effective number of layer is smaller by 2 than $N_\mathrm{L}$, which simply means that the system is slightly less correlated than if two segments in contact had exactly the same length. This robust behavior shows that the statistics is completly determined by the average number of layers in the ball.

\begin{figure}
    \centering
    \includegraphics[width=.95\columnwidth]{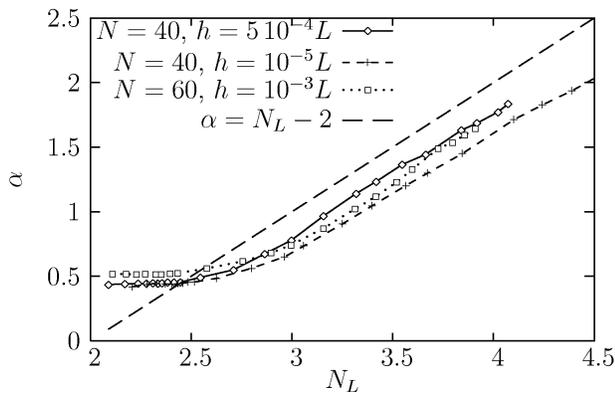}
    \caption{The parameter $\alpha$ of  the Gamma distribution $P_\Gamma$ (Eq.~\ref{pg}) as a function of the number of layers $N_\mathrm{L}$ (defined as the mean number of intersections of a half-infinite straight line with the system, averaged over all directions of the line and all samples).}
    \label{fig4}
\end{figure}

To summarise, we have introduced a simple 1D model which allows a comprehensive statistical study of crumpled paper. At low confinement, a hierarchy of buckling events leads to a log-normal distribution of lengths in agreement with the predictions of~\cite{witten} and the experiments of~\cite{blair}. At high confinement, jamming is approached with a critical behavior; self-avoidance leads to self-correlations and to a Gamma distribution of lengths solely determined by the number of layers in the system. This increase in correlations accounts for the Hurst exponent of compact paper balls~\cite{ploura} being larger than in the loose balls of~\cite{blair}. Obviously, our results call for experimental measurements of length distributions at high confinement. Besides, our model can be refined in two ways. First, the number of vertices could be made free but one needs to find a criterion to add or suppress vertices. Second, it could be extended to 2D which is more realistic but yields a more difficult numerical task : if the sheet is assumed to be a polyhedron, developability becomes a local constraint at each vertex whereas self-avoidance becomes even more computationally expensive.

This study was partially supported by the Minist\`ere de la Recherche--ACI Jeunes Chercheurs. We are grateful to Mokhtar Adda-Bedia, Etienne Couturier and St\'ephane Douady for discussions and to E. C. for Fig.~1a.

\bibliography{crumpled_pap}

\end{document}